\documentclass[twocolumn,amsmath,amssymb,aps,physrev,twocolumn,superscriptaddress]{revtex4-2}
\usepackage[utf8]{inputenc}
\usepackage{float}
\usepackage{booktabs}
\usepackage{amsmath}
\usepackage{graphicx}

\makeatletter

\providecommand{\tabularnewline}{\\}

\usepackage{graphicx}
\usepackage{dcolumn}
\usepackage{bm}

\makeatother

\begin{document}
\preprint{APS/123-QED}
\title{Wave-number-dependent closure condition for fluid moment equations}
\author{Yong Sun}
\affiliation{Advanced Energy Science and Technology Guangdong Laboratory, Huizhou
516000, China}
\affiliation{Institute of Modern Physics, Chinese Academy of Sciences, Lanzhou
730000, China}
\author{Shijia Chen}
\affiliation{Center for Intense Laser Application Technology, and College of Engineering
Physics, Shenzhen Technology University, Shenzhen 518118, China}
\author{Minqing He}
\affiliation{Institute of Applied Physics and Computational Mathematics, Beijing
100094, China}
\author{Sizhong Wu}
\affiliation{Center for Intense Laser Application Technology, and College of Engineering
Physics, Shenzhen Technology University, Shenzhen 518118, China}
\author{Rui Cheng}
\author{Jie Yang}
\affiliation{Advanced Energy Science and Technology Guangdong Laboratory, Huizhou
516000, China}
\affiliation{Institute of Modern Physics, Chinese Academy of Sciences, Lanzhou
730000, China}
\affiliation{State Key Laboratory of Heavy Ion Science and Technology, Institute
of Modern Physics, Chinese Academy of Sciences, Lanzhou 730000, China}
\author{Lei Yang}
\author{Zhiyu Sun}
\author{Liangwen Chen}
\email{chenlw@impcas.ac.cn}

\affiliation{Advanced Energy Science and Technology Guangdong Laboratory, Huizhou
516000, China}
\affiliation{Institute of Modern Physics, Chinese Academy of Sciences, Lanzhou
730000, China}
\affiliation{State Key Laboratory of Heavy Ion Science and Technology, Institute
of Modern Physics, Chinese Academy of Sciences, Lanzhou 730000, China}
\affiliation{School of Nuclear Science and Technology, University of Chinese Academy
of Sciences, Beijing 100049, China}
\author{Hua Zhang}
\email{zhanghua@sztu.edu.cn}

\affiliation{Center for Intense Laser Application Technology, and College of Engineering
Physics, Shenzhen Technology University, Shenzhen 518118, China}
\begin{abstract}
Fluid models offer crucial computational efficiency for plasma simulations,
yet accurately capturing kinetic effects like Landau damping remains
a fundamental challenge. While conventional closures (e.g., Hammett-Perkins
and Hunana) are widely used, their fidelity relative to exact kinetic
response degrades significantly depending on the perturbation wave
number. Here, we propose a novel wave-number-dependent closure condition
for the three-moment fluid equations that explicitly preserves the
primary dispersion relation. By mapping Padé approximant coefficients
directly to the kinetic roots of the collisionless Vlasov-Poisson
system, we derive an analytical closure that rigorously embeds exact
kinetic scaling across all spatial scales. We further demonstrate
that this framework readily extends to collisional plasmas via the
BGK model. This deterministic approach precisely captures the long-term
macroscopic evolution of fluid moments and field energy, offering
a rigorous foundation for high-fidelity fluid modeling. 
\end{abstract}
\maketitle

Fluid models offer crucial computational efficiency for plasma simulations,
but accurately capturing kinetic effects like Landau damping remains
a fundamental challenge \citep{snyderLandauFluidModel2001}. The fidelity
of these models hinges entirely on the closure condition used to truncate
the infinite hierarchy of moment equations derived from kinetic theory
\citep{chenIntroductionPlasmaPhysics2015}. While early local closures
such as the Spitzer--Härm \citep{spitzerTransportPhenomenaCompletely1953}
and Braginskii \citep{braginskiiTransportProcessesPlasma1965} models
lacked kinetic effects, the Hammett-Perkins (HP) closure revolutionized
fluid modeling by introducing a non-local transport relation to capture
wave-particle resonance \citep{hammettFluidMomentModels1990}. This
foundational approach was later refined by Hunana et al., who utilized
Padé approximants to more accurately model the plasma response function
\citep{hunanaNewClosuresMore2018}.

Despite these advancements, a critical limitation of existing closures
is their static nature. Because their coefficients are derived purely
from asymptotic limits, their accuracy degrades significantly at intermediate
scales. Recently, data-driven approaches, such as Fourier neural operators,
have been deployed to learn implicit, non-local heat flux mappings
from fully kinetic data to reproduce Landau damping more accurately
\citep{huangMachinelearningHeatFlux2025,wangDeepLearningSurrogate2020,maMachineLearningSurrogate2020}.
While these machine learning surrogate models successfully capture
macroscopic plasma evolution, their theoretical mechanisms are less
transparent and they introduce computational overhead inherent to
training neural networks.

A more deterministic, analytical pathway that bridges this gap is
highly desirable. In this Letter, we address the problem by proposing
a novel, wave-number-dependent closure condition for the three-moment
fluid equations. Instead of relying on static asymptotic coefficients,
our framework explicitly anchors the fluid closure to the physical
dispersion relation. By mapping Padé approximant coefficients directly
to the kinetic roots of the Vlasov-Poisson system, we derive an analytical
closure that dynamically adapts to specific wave numbers, preserving
exact kinetic effects across all scales.

To establish the core physics, we first consider the one-dimensional,
collisionless motion of electrons in a plasma governed by the Vlasov
equation:

\begin{equation}
\frac{\partial f}{\partial t}+v\frac{\partial f}{\partial z}+\frac{F}{m}\frac{\partial f}{\partial v}=0,
\end{equation}
where $f(z,v,t)$ is the electron velocity distribution function at
spatial coordinate $z$, velocity $v$, and time $t$. The force term
$F=-eE=e\frac{\partial\phi}{\partial z}$ arises from the self-consistent
electric field $E$, with $\phi$ being the electric potential, $e$
the elementary charge, and $m$ the electron mass. Taking the velocity
moments of the Vlasov equation, specifically the zeroth moment (density
$n$), first moment (velocity $u$), and second moment (pressure $P$),
yields the standard hierarchy of fluid evolution equations. This macroscopic
system is fundamentally unclosed because the pressure evolution depends
on the unknown third moment, the heat flux $q$:
\begin{equation}
\begin{aligned}\frac{\partial n}{\partial t}+\frac{\partial(nu)}{\partial z}=0,\\
\frac{\partial u}{\partial t}+u\frac{\partial u}{\partial z}+\frac{1}{mn}\frac{\partial P}{\partial z}+\frac{e}{m}E=0,\\
\frac{\partial P}{\partial t}+u\frac{\partial P}{\partial z}+3P\frac{\partial u}{\partial z}+\frac{\partial q}{\partial z}=0.
\end{aligned}
\end{equation}

To determine an expression for $q$, we analyze the linear response
of an initially homogeneous Maxwellian plasma to a small driving perturbation
proportional to $\exp(ikz-i\omega t)$, where $k$ is the perturbation
wave number and $\omega$ is the frequency. In Fourier space, the
first-order density perturbation $n_{1k}$ relates to the perturbed
potential $\phi_{1k}$ via:

\begin{equation}
n_{1k}=\frac{e\phi_{1k}n_{0}}{T_{0}}R(\zeta),
\end{equation}
where $n_{0}$ is the equilibrium density, $T_{0}=mv_{t}^{2}$ is
the initial temperature ($v_{t}$ being the thermal velocity), $R(\zeta)$
is the kinetic response function, and $\zeta=\omega/(\sqrt{2}v_{t}|k|)$
is the normalized phase speed. Following the pioneer work of Hunana
et al. \citep{hunanaNewClosuresMore2018,hunanaIntroductoryGuideFluid2019},
for a three-moment fluid model, we approximate $R(\zeta)$ using a
$3$-pole Padé approximant: 
\begin{equation}
R(\zeta)\approx R_{3,0}(\zeta)\equiv\frac{1+a_{1}\zeta}{1+b_{1}\zeta+b_{2}\zeta^{2}-2a_{1}\zeta^{3}}.
\end{equation}
Combining the linearized moment equations yields a general closure
condition for the heat flux:

\begin{equation}
\hat{q}=Q_{1}\hat{u}+iQ_{2}\hat{k}(\hat{p}-\hat{\phi})-iQ_{3}\hat{k}\hat{T},
\end{equation}
where $\hat{k}\equiv k/|k|$ is the direction of the wave vector.
The normalized perturbed variables are defined as $\hat{n}=n_{1k}/n_{0}$,
$\hat{u}=u_{1k}/(\sqrt{2}v_{t})$, $\hat{p}=P_{1k}/P_{0}$, $\hat{q}=q_{1k}/(P_{0}\sqrt{2}v_{t})$,
$\hat{\phi}=e\phi_{1k}/T_{0}$, and the temperature perturbation $\hat{T}=\hat{p}-\hat{n}$.
The closure coefficients $Q_{1}$, $Q_{2}$, and $Q_{3}$ are strictly
determined by the chosen Padé approximant, or more concretely, $Q_{1}=\frac{b_{1}-3a_{1}}{a_{1}},\ Q_{2}=\frac{1+b_{2}/2}{ia_{1}},\ Q_{3}=\frac{1}{ia_{1}}.$
The Padé coefficients are then determined by matching the Padé form
with the asymptotic power series of $R(\zeta)$. Matching different
orders of $\zeta$ from the adiabatic ($|\zeta|\ll1$) and fluid limit
($|\zeta|\gg1$) leads to different values of Padé coefficients, as
summarized in Table 1.
\begin{table}[H]
\centering
\begin{centering}
\begin{tabular}{cccc}
\toprule 
Approximant & $a_{1}$ & $b_{1}$ & $b_{2}$\tabularnewline
\midrule
$R_{3,0}$ & $-i\sqrt{\pi}\frac{\pi-3}{4-\pi}$ & $-i\frac{\sqrt{\pi}}{4-\pi}$ & $-\frac{3\pi-8}{4-\pi}$\tabularnewline
$R_{3,1}$ & $-i\frac{4-\pi}{\sqrt{\pi}}$ & $\frac{-4i}{\sqrt{\pi}}$ & $-2$\tabularnewline
$R_{3,2}$ (HP) & $\frac{-i\sqrt{\pi}}{2}$ & $\frac{-3i\sqrt{\pi}}{2}$ & $-2$\tabularnewline
\bottomrule
\end{tabular}
\par\end{centering}
\caption{Values of Padé coefficients from asymptotic form. The $i$ indices
in $R_{3,i}$ indicate the orders of $\zeta$ matched from the fluid
limit, $R_{3,0}$ being the leading order and so on. $R_{3,2}$ leads
to HP closure.}
\end{table}

A critical oversight in existing asymptotic derivations is that they
determine these Padé coefficients without accounting for the influence
of the electric field on the heat flux via the Poisson equation, $k^{2}\phi_{1k}=-en_{1k}$.
Substituting the Poisson relation into the density perturbation imposes
the exact kinetic dispersion relation: 
\begin{equation}
R(\zeta)+k^{2}/k_{p}^{2}=0,
\end{equation}
where $k_{p}\equiv e\sqrt{n_{0}/T_{0}}$ is the Debye wave number.
All roots of this equation possess negative imaginary parts, dictating
the temporal decay (Landau damping) of the perturbations. The roots
with imaginary parts closest to zero (the least-damped roots) dominate
the long-term physical behavior.

To enforce these exact kinetic characteristics within the fluid model,
our framework requires that the Padé approximant explicitly shares
these least-damped kinetic roots $\zeta_{j}(k^{2})$. Substituting
these roots into the approximated dispersion relation provides a solvable
system of equations. This allows us to determine the Padé coefficients,
and consequently $Q_{1}$, $Q_{2}$, and $Q_{3}$, as explicit functions
of $k^{2}$, dynamically adapting the fluid closure to the perturbation
wave number. 

For plasmas with parity symmetry, specifically our collisionless model
with initially Maxwellian distributions, the kinetic roots are symmetric
about the imaginary axis. For demonstrative purposes, we would utilize
the pair of least-damped roots $\zeta_{0}$ and $-\zeta_{0}^{*}$
and the $R_{3,1}$ approximant. The combination automatically ensures
that the solutions of $a_{1},b_{1}$ are purely imaginary and that
the $Q_{i}'s$ are real. One may also utilize the $R_{3,0}$ approximant
but that would require a pseudo-root on the imaginary axis (e.g. $i\mathrm{Im(\zeta_{1})}$)
to maintain the overall symmetry, which introduces an artificial error
and is generally not desirable.  
\begin{figure}[t]
\centering
\includegraphics[scale=0.9]{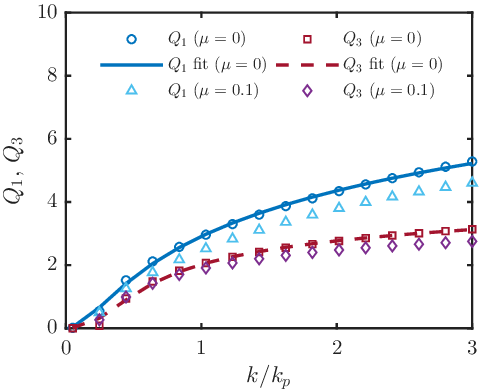}

\caption{Fitting of the Closure Parameters. $Q_{i}$ data points are solved
exactly using the root-matching ansatz. The fitted functions are $Q_{1}(\tilde{k})\approx1.08\ln(1+13.73\tilde{k}^{2})$
and $Q_{3}(\tilde{k})\approx1.14\sqrt{\ln(1+21.92\tilde{k}^{4})}$
for the collisionless ($\mu=0$) case with $\tilde{k}\equiv k/k_{p}$.
The closure parameters for weakly collisional BGK model ($\mu=0.1$)
is also plotted for reference.}
\end{figure}

Since $b_{2}=-2$ for the $R_{3,1}$ approximant, $Q_{2}=0$. To determine
the fitting functions for $Q_{1}$ and $Q_{3}$, we consider their
asymptotic behavior: as $k\rightarrow0$, $Q_{1}\sim k^{2}$ and $Q_{3}\sim\exp(-k^{-2})$;
as $k\rightarrow\infty$, $Q_{1}\sim\ln k$ and $Q_{3}\sim\sqrt{\ln k}$.
An educated guess leads to $Q_{1}(k)=c_{1}\ln(1+c_{2}k^{2})$ and
$Q_{3}(k)=d_{1}\sqrt{\ln(1+d_{2}k^{4})}$, which achieves a balance
of accuracy and elegance.  Fig. 1 demonstrates the fitting process
and provides an explicit expression for the collisionless closure
parameters.

\begin{figure}[t]
\centering
\includegraphics[scale=0.9]{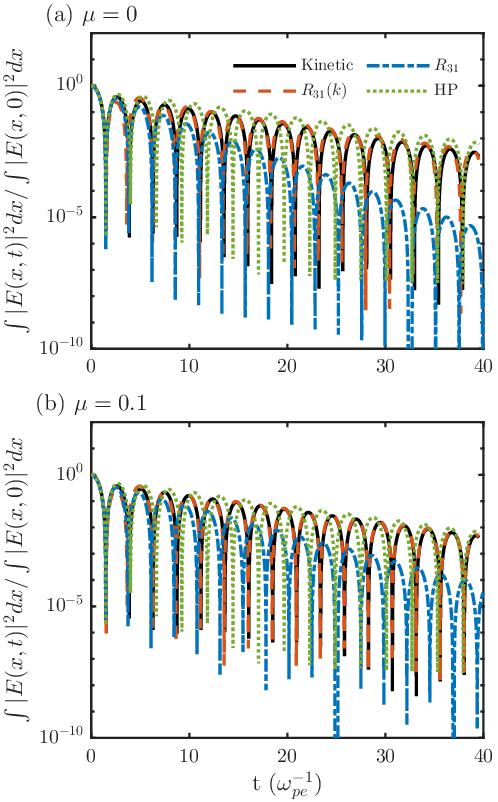}

\caption{Evolution of normalized electric field energy. Comparison of different
closure conditions ($k$-dependent closure, red dashed line; HP closure,
green dotted line; Hunana $R_{3,1}$ closure, blue dash-dotted line)
benchmarked against a Vlasov simulation (black solid line), with $k=0.4k_{p}$.
Top panel is for collisionless case, while bottom panel is collisional
($\mu=0.1$).}
\end{figure}

To verify this ansatz, we benchmark our wave-number-dependent fluid
closures against fully kinetic 1D-1V Vlasov-Poisson simulations. The
fluid moment equations are evolved in Fourier space using a second-order
Runge-Kutta scheme, while the kinetic benchmark utilizes a Fourier
spectral method with operator splitting. We initialize the system
with a Maxwellian distribution perturbed by a small density fluctuation
$f(z,t=0)=f_{0}(1+A\cos(kz))$ (amplitude $A=0.02$, wave number $k=0.4k_{p}$).
The spatial domain is periodic, with $0<z\le2\pi/k$ evenly divided
into 128 segments. The time step is set as $0.005\omega_{pe}^{-1}$
and we evolve the fluid system until $40\omega_{pe}^{-1}$. For the
kinetic simulation, the velocity space ranges from $-8v_{t}$ to $8v_{t}$,
evenly divided into 256 segments, the Runge-Kutta is set to third
order to ensure good precision.

The normalized electric field energy $\int|E(z,t)|^{2}dx$ is shown
in Fig. 2a. Conventional closures ($R_{3,1}$ and HP) with constant
Padé coefficients roughly capture the initial Landau damping, but
their envelope curves deviate rapidly from the kinetic benchmark.
In stark contrast, our wave-number-dependent closure accurately tracks
the kinetic result, nearly perfectly reproducing the exact long-term
Landau damping behavior.

To demonstrate the robustness and applicability of this ansatz, we
extend the collisionless model to incorporate collisions. Dissipation
term is introduced by replacing the zero on the right-hand side of
Eq. (1) with the BGK operator $-\mu(f-f_{eq})$, where $\mu$ is the
collision frequency and $f_{eq}$ is the local Maxwellian \citep{bhatnagarModelCollisionProcesses1954a}.
Because the BGK operator conserves the primary moments, the fluid
equations (Eq. 2) remain structurally identical. However, in the weak
collision limit, the kinetic dispersion relation modifies to
\begin{align}
 & 1+KR(\zeta_{c})\nonumber \\
 & +i\nu\left[\left(2\zeta_{c}^{3}-K\zeta_{c}\right)R(\zeta_{c})+\frac{3+K}{2}\frac{R(\zeta_{c})-1}{\zeta_{c}}+\zeta_{c}\right] & =0,
\end{align}
where $K=k_{p}^{2}/k^{2}$, $\zeta_{c}=\zeta+i\nu$ and $\nu=\frac{\mu}{\sqrt{2}|k|v_{t}}$.
The general form of the closure from $R_{3,0}$ approximant is determined
as:
\begin{align}
\hat{q}_{BGK} & =[Q_{1}-\nu Q_{2}(4+Q_{1})]\hat{u}+[Q_{2}-\nu Q_{2}^{2}]i\hat{k}(\hat{p}-\hat{\phi})\nonumber \\
 & +[\nu(Q_{1}/2+Q_{2}Q_{3})-Q_{3}]i\hat{k}(\hat{p}-\hat{n})
\end{align}
In the collisionless limit $\mu\rightarrow0$, both the kinetic dispersion
relation and the form of closure restores to the collisionless case
presented in Eqs. (5) and (6). Note that parity symmetry is still
conserved in this model. Similarly, we may solve for the pair of least-damped
roots and use them to determine the k-dependent Padé coefficients
for $R_{3,1}$ approximant (where $Q_{2}=0$). The resulting closure
parameters now depends on both the wave number $k$ and collision
frequency $\mu$. Specifically for $\mu=0.1$, we have $Q_{1}(\tilde{k})\approx0.98\ln(1+11.80\tilde{k}^{2})$
and $Q_{3}(\tilde{k})\approx0.97\sqrt{\ln(1+42.23\tilde{k}^{4})}$.
The new closure for BGK model is also compared with the conventional
ones and benchmarked against the kinetic result. As visualized in
Fig. 2, our new closure behaves consistently better for both the collisional
and collisionless cases.

\begin{figure*}[t]
\centering
\includegraphics[scale=0.85]{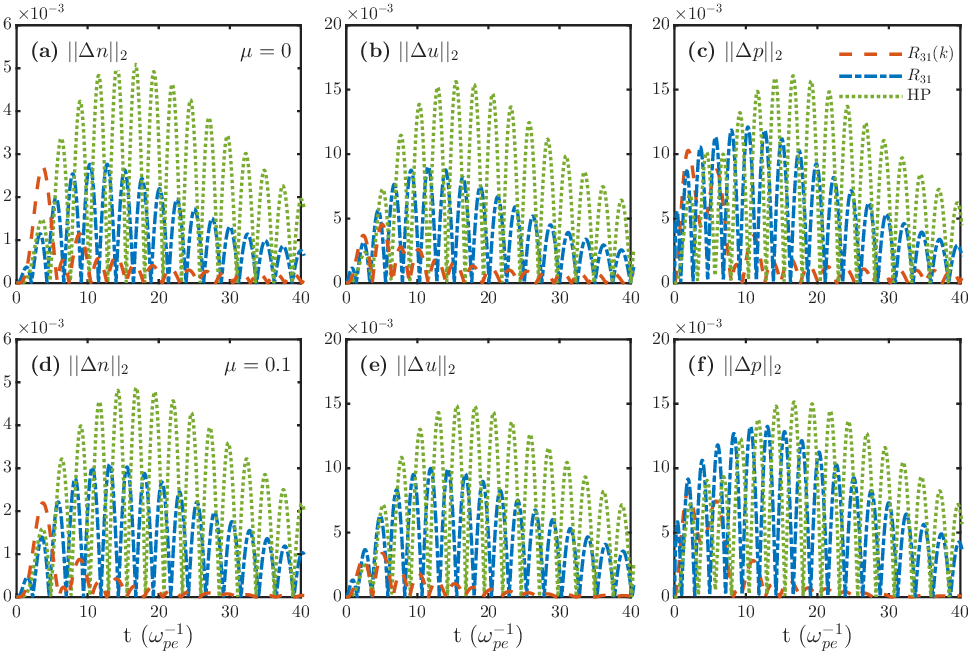}

\caption{Deviations of fluid moments from kinetic results. Spatial L2-norm
errors of lower-order moments using different closure conditions ($k$-dependent
closure, red dashed line; HP closure, green dotted line; Hunana $R_{3,1}$
closure, blue dash-dotted line) compared to kinetic benchmarks, with
$k=0.4k_{p}$. Subplots (a)-(c) depict the collisionless case ($\mu=0$),
while (d)-(f) depict the collisional case ($\mu=0.1$).}
\end{figure*}

Fig. 3 further plots the spatial L2-Norm error of lower-order moments
via different closure conditions compared with the kinetic result.
It is evident that as time elapses, the wave-number-dependent closure
produces significantly (around $10\times$) smaller errors compared
to the conventional closure.

The underlying kinetic dispersion relation dictates long-term macroscopic
plasma behavior, and its active preservation represents a significant
advancement in constructing high-fidelity fluid closures. As demonstrated,
dynamically mapping fluid coefficients to the least-damped kinetic
roots enables the model to accurately reproduce the long-term evolution
of field energy and macroscopic moments. This approach offers substantial
improvements over conventional static closures in both collisionless
and BGK collisional models.

Furthermore, while the current ansatz utilizes the parity symmetry
of initially Maxwellian plasmas to satisfy the Hermitian constraint
of a physical heat flux, this framework can be readily generalized.
For asymmetric systems, such as drift-wave dynamics in strongly magnetized
plasmas \citep{hortonDriftWavesTransport1999}, extending to a 4-moment
model with a higher-order Padé approximant $(R_{4,2})$ provides a
viable path forward. Such a structural expansion offers the necessary
mathematical degrees of freedom to accurately map asymmetric least-damped
roots while rigorously ensuring that macroscopic observables remain
strictly real.

Ultimately, by anchoring macroscopic fluid closure directly to the
fundamental dispersion relation, this wave-number-dependent framework
resolves a long-standing theoretical bottleneck. It provides a robust,
analytically transparent pathway to model multiscale kinetic phenomena,
paving the way for more physically accurate studies of turbulence,
turbulent transport, and macroscopic instabilities in advanced plasma
systems

\begin{acknowledgments}
This work is supported by the Dongjiang Laboratory S\&T Program (DJL2025C009),
the Research Program of State Key Laboratory of Heavy Ion Science
and Technology, Institute of Modern Physics, Chinese Academy of Sciences,
under Grant No. HIST2025CS06, the National Key R\&D Program of China
(Grant No. 2022YFA1603302), the National Natural Science Foundation
of China (Grant No. 12105327, 12375237, 12305275), and the Guangdong
Basic and Applied Basic Research Foundation (Grant No. 2023B1515120067).
\end{acknowledgments}

\appendix
\bibliographystyle{apsrev4-2}
\bibliography{closure}

\end{document}